\documentclass[11pt]{article}

\usepackage{amsmath}
\usepackage{amsfonts}
\usepackage{latexsym}
\begin{document}
\title{{\huge A note on wavemap-tensor cosmologies\thanks{Based on a talk presented by the second
author at the 2nd Hellenic Cosmology Workshop,
National Observatory of Athens, April 21-22, 2001.}}}
\author{{\Large Spiros Cotsakis}\\
GEODYSYC\\
Department of Mathematics\\
University of the Aegean\\
Karlovassi 83 200, Greece\\
\texttt{email: skot@aegean.gr}\\
\and
{\Large John Miritzis}\\
Department of Marine Sciences\\
University of the Aegean\\
5, Sapfous Str., Mytilene 81 100, \\ Greece\\
\texttt{email: john@env.aegean.gr}}
\maketitle
\begin{abstract}
\noindent  We examine theories of gravity which include finitely many coupled
scalar fields with arbitrary couplings to the curvature
(wavemaps). We show that the most general scalar-tensor
$\sigma$-model action is conformally equivalent to general
relativity with a minimally coupled wavemap with a particular
target metric. Inflation on the source manifold is then shown to
occur in a novel way due  to the combined effect of arbitrary
curvature couplings and  wavemap self-interactions. A new
interpretation of the conformal equivalence theorem proved for
such `wavemap-tensor' theories  through brane-bulk dynamics is
also discussed.
\end{abstract}

\section{Introduction}
Scalar fields currently play a prominent role in the construction
of cosmological scenarios aiming at describing the structure and
evolution of the early universe. The standard inflationary idea
requires that there is a period of slow-roll evolution of a scalar
field (the inflaton) during which its potential energy dominates
the kinetic energy and drives the universe in a quasi-exponential,
accelerated expansion \cite{ko-tu90,li91}. In string cosmology,
there is another rolling scalar field (the dilaton) whose
\emph{kinetic} energy drives a stage of accelerated contraction
\cite{ga-ve93}. The main dynamical role of the existence of scalar
fields in such contexts lies in driving periods of accelerated
evolution of the scale factor. During such phases of evolution the
curvature grows (as in a string or a Kaluza-Klein general
relativistic  context), or is constant or slightly decreasing (as
in standard inflationary dynamics) leading through the behaviour
of the Hubble radius $|H|^{-1}$ to shrinking or growing horizons.
The close interplay of the time behaviours of scale factors and
horizon scales in such models leads to fundamental properties
characterizing such cosmologies and to important current ideas for
the behaviour of spacetime near the Planck scale \cite{ga00}.

Scalar fields also arise naturally in  alternative theories of
gravity which aim at extending general relativity, e.g.,  higher
order gravity theories,   scalar-tensor and string theories. In
higher derivative gravity, due to its conformal relation with
general relativity \cite{ba-co88}, scalar fields appear in the
Einstein frame with a self-interaction, nonlinear  potential term
which mimics the higher order curvature properties of the original
(Jordan) frame. In a scalar-tensor theory, there are scalar fields
which are typically coupled nonminimally to the curvature leading
to interesting nonsingular cosmologies even in the isotropic
category \cite{ba93}. (The simplest scalar-tensor theory developed
by Brans and Dicke \cite{br-di61} involves a massless scalar field
with constant coupling to matter. Generalizations of the
Brans-Dicke theory lead to  scalar-tensor theories with scalar
field self-interactions and dynamical couplings to matter. Further
generalizations can be achieved by considering multiple scalar
fields (see e.g., \cite{da-es92})). Similar results can be proved
for simple bosonic string theories \cite{cfmp85}.

All problems above may be considered in the Einstein frame for an
Einstein-scalarfield system defined on spacetime for we know that
due to their conformal properties all couplings of a scalar field
to the curvature are equivalent \cite{pa91}. In this
paper  we
prove, among other things, a scalar-tensor generalization of this
result for any self-interacting system of a finite number of
scalar fields. Such objects are known as \emph{wavemaps} to
mathematicians \cite{ch99,sh-st98} and as \emph{nonlinear-$\sigma$
models} to physicists. We shall also see that brane cosmologies
or, in general, the classical dynamics of `brane objects' (see,
for example, \cite{ca00}) can be obviously described through (in
fact they are completely equivalent to) wavemaps.

The plan of this paper is as follows. In the next Section, we
present the basic equations of our theory. Section 3 proves the
conformal equivalence theorem for our wavemap-tensor theory to an
Einstein-wavemap system thus showing the generalization that all
couplings of a wavemap to the curvature are equivalent. In Section
4, we discuss a new way to implement an inflationary phase in this
theory through a mechanism completely distinct from others that
have appeared in the literature and show that the theory allows in
a natural way to built a cosmological constant as remnant of the
early state of the universe.  We conclude in Section 5 giving some
connections of the present set of problems with those of classical
brane dynamics.

\section{Wavemaps and multiscalar-tensor gravity theories}
A wavemap is a map from a spacetime manifold to any (semi-)
Riemannian manifold. More precisely, consider a spacetime
$(\mathcal{M}^{m},g_{\mu\nu}),$ a Riemannian manifold
$(\mathcal{N}^{n},h_{ab})$ and a $\mathcal{C}^{\infty}$ map
$\phi:\mathcal{M}\rightarrow\mathcal{N}$. We call $\mathcal{M}$
the source manifold and $\mathcal{N}$ the target
manifold\footnote{We use greek indices for tensorfields defined on
the source manifold $\mathcal{M}$  and latin indices for those on
the target manifold $\mathcal{N}.$ $\Gamma_{\mu\nu}^{\rho}$
denotes the metric connection of the spacetime $\mathcal{M}$
while $\Gamma_{bc}^{a}$ is the metric connection of the target
manifold $\mathcal{N}.$}. For instance, if
$\mathcal{N}=\mathbb{R}$ then $\phi$ is simply a real scalar field
on the source $\mathcal{M}.$ If, on the other hand, we choose
$\mathcal{N}=\mathbb{R}^{n}$ then the wavemap
$\phi=\left(\phi^{1},...,\phi^{n}\right)$ may be thought of as $n$
\emph{uncoupled} scalar fields on $\mathcal{M}.$ We may think of
the scalar fields $\phi^{a},\; a=1,\dots,n$, as coordinates
parametrizing the Riemannian target. The metric $h_{ab}$ (which in
general is not flat) expresses the possible couplings of the
scalar fields.

A $\mathcal{C}^{\infty}$ map
$\phi:\mathcal{M}\rightarrow\mathcal{N}$ as above is called
\emph{a wavemap} if it is a critical point of the action,
\begin{equation}
S=\int_{\mathcal{M}}L\; dv_{g},\;\;dv_{g}=\sqrt{-g}dx,
\end{equation}
with,
\begin{equation}
L=-g^{\mu\nu}h_{ab}\partial_{\mu}\phi^{a}\partial_{\nu}\phi^{b}.\label{wplagr}
\end{equation}
The Euler-Lagrange equations for this action are,
\begin{equation}
\Box_{g}\phi^{a}+\Gamma_{bc}^{a}(h)g^{\mu\nu}\partial_{\mu}\phi^{b}%
\partial_{\nu}\phi^{c}=0,
\end{equation}
and constitute a quasi-linear system of hyperbolic PDEs for
$\phi^{a}.$

We can consider a wavemap coupled to the curvature that is, regard
wavemaps as `sources' of the gravitational field and start with
the Hilbert action,
\begin{equation}
S=\int_{\mathcal{M}}\left(
R-g^{\mu\nu}h_{ab}\partial_{\mu}\phi^{a}
\partial_{\nu}\phi^{b}\right)  dv_{g}, \label{eiwmaction}
\end{equation}
where $R$ is the scalar curvature of the source $\mathcal{M}$.
Varying this action with respect to the metric $g$ and the fields
$\phi^{a}$ we arrive at the so-called Einstein-wavemap system,
namely,
\begin{eqnarray}
G_{\mu\nu}&=&h_{ab}\left(
\partial_{\mu}\phi^{a}\partial_{\nu}\phi^{b}-\frac
{1}{2}g_{\mu\nu}g^{\rho\sigma}\partial_{\rho}\phi^{a}\partial_{\sigma}\phi
^{b}\right) ,\label{ewp1a}\\
\Box_{g}\phi^{a}&+&\Gamma_{bc}^{a}(h)g^{\mu\nu}\partial_{\mu}\phi^{b}%
\partial_{\nu}\phi^{c}=0\; .\label{ewp1b}
\end{eqnarray}
The stress-energy tensor of the wavemap is defined in the standard
way through the basic wavemap lagrangian (\ref{wplagr}) and is
given by,
\begin{equation}
T_{\mu\nu}=h_{ab}\left(
\partial_{\mu}\phi^{a}\partial_{\nu}\phi^{b}-\frac
{1}{2}g_{\mu\nu}g^{\rho\sigma}\partial_{\rho}\phi^{a}\partial_{\sigma}\phi
^{b}\right)
\end{equation}
This has the nice properties of being a symmetric, divergence-free
tensorfield and satisfies $T_{\mu \nu}u^{\mu}u^{\nu}\geq0$, for
all future-directed timelike vector fields $u^{\mu}$ on the sourse
spacetime. In the simple case where $\mathcal{N}=\mathbb{R},$ we
see that $T_{\mu\nu}$ is reduced to the stress-energy tensor of a
massless scalar field.

Our starting point is the general scalar-tensor action functional,
\begin{equation}
S=\int_{\mathcal{M}}L_{\sigma}dv_{g},\;\;dv_{g}=\sqrt{-g}dx,\label{action}%
\end{equation}
with,
\begin{equation}
L_{\sigma}=A(\phi)R-B(\phi)g^{\mu\nu}h_{ab}\partial_{\mu}\phi^{a}\partial
_{\nu}\phi^{b},\label{lagran}
\end{equation}
where $A,B$ are arbitrary $\mathcal{C}^{\infty}\;$ functions of
$\phi$. This class of gravity theories which we call
\emph{wavemap-tensor theories}, includes as special cases many of
the scalar field models considered in the literature e.g.,
\cite{da-es92,barr,stac}. Choosing $A\left(\phi\right) =\phi$ and
$B\left( \phi\right) =\omega/\phi,$ with $\omega=const.,$ we
recognize the standard Brans-Dicke theory. Secondly, taking
$\mathcal{N}=\mathbb{R},$ with $A\left( \phi\right)
=B\left(\phi\right)  =1,$ we obtain General Relativity with a
massless scalar field as the matter source. In the case of an
arbitrary Riemannian manifold $\mathcal{N},$ setting
$A\left(\phi\right) =B\left(\phi\right) =1,$ the field equations
derived upon variation of the corresponding action (\ref{action})
reduce to the Einstein-wavemap system (\ref{ewp1a},\ref{ewp1b}).

In the general case,  $S$ as given in
(\ref{action})-(\ref{lagran}) has arbitrary couplings to the
curvature and kinetic terms.  Varying it with respect to the
metric $g$ and the scalar fields $\phi\;$ we obtain the system,
\begin{align}
G_{\mu\nu}
&
=\frac{B}{A}h_{ab}\left(  \partial_{\mu}\phi^{a}\partial_{\nu
}\phi^{b}-\frac{1}{2}g_{\mu\nu}g^{\rho\sigma}\partial_{\rho}\phi^{a}%
\partial_{\sigma}\phi^{b}\right)
\nonumber\\
&
+\frac{1}{A}\left(  \nabla_{\mu}\nabla_{\nu}A-g_{\mu\nu}\Box_{g}A\right)  ,
\label{field1}%
\end{align}%

\begin{equation}
\Box_{g}\phi^{a}+\bar{\Gamma}_{bc}^{a}g^{\mu\nu}\partial_{\mu}\phi^{b}%
\partial_{\nu}\phi^{c}+\frac{1}{2}RA^{a}=0,\quad\bar{\Gamma}_{bc}^{a}%
=\Gamma_{bc}^{a}(h)+C_{bc}^{a},\label{wmap1}%
\end{equation}
where we have set $A_{a}=\partial A/\partial\phi^{a},$
$C_{bc}^{a}=(1/2)\left(\delta_{b}^{a}B_{c}+\delta_{c}^{a}
B_{b}-h_{bc}B^{a}\right)$ and $B_{a}=\partial\ln
B/\partial\phi^{a}$.

We see that $\phi$ satisfies a wavemap-type equation with the
connection coefficients $\bar{\Gamma}_{bc}^{a}$ defining a Weyl
geometry in $\mathcal{N}.$ \footnote{However, the Weyl vector
$B_{a}$ is a gradient and can be gauged away by the conformal
transformation of the target metric, $\tilde {h}_{ab}=B\left(
\phi\right) h_{ab}$ (see, for example, Schouten \cite{scho}). We
find $\bar{\Gamma}_{bc}^{a}=\Gamma_{bc}^{a}(\tilde{h})$ that is,
the connection is the Levi-Civita connection of the metric
$\tilde{h}$. This result is already clear from the form of the
general wavemap-tensor action and we could have absorbed from
the beginning the function $B\left(\phi\right)$ into the metric of
the target manifold.} In the following, without loss of
generality, we  set $B=1$ in Eq. (\ref{lagran}) and drop the tilde
on $h$.

\section{Conformal structure of wavemap-tensor theories}

The right-hand side of the field equation (\ref{field1}), defines
an `energy-momentum tensor' and splits into two parts. The first
term, apart from a proportionality function, is exactly the
energy-momentum tensor of the wavemap, namely,
\begin{equation}
h_{ab}\left(  \partial_{\mu}\phi^{a}\partial_{\nu}\phi^{b}-\frac{1}{2}%
g_{\mu\nu}g^{\rho\sigma}\partial_{\rho}\phi^{a}\partial_{\sigma}\phi
^{b}\right)  .
\end{equation}
The second part is the well-known combination,
\begin{equation}
\nabla_{\mu}\nabla_{\nu}A-g_{\mu\nu}\Box_{g}A,
\end{equation}
which contains second order covariant derivatives of the fields.
Therefore the \emph{total} energy-momentum tensor does not
necessarily satisfy the strong energy condition, i.e. the energy
density of the fields cannot always be made  nonnegative.
Furthermore,
it contains terms proportional to the connection $\Gamma_{\mu\nu}%
^{\rho}$ responsible for the dynamical evolution of the
gravitational field and therefore we could ascribe to these terms
a physical meaning as properly  `belonging' to the LHS of the
field equations (\ref{field1}) and not be part of the `material
content' of the theory.

These difficulties as well as many others may be overcome by
performing a suitable conformal transformation on the source
manifold and redefining the fields in suitable ways. Defining a
new metric by,
\begin{equation}
\tilde{g}_{\mu\nu}=A(\phi)g_{\mu\nu},
\end{equation}
the scalar curvature transforms as,
\begin{equation}
R=A\left(  \tilde{R}+3\tilde{\Box}_{\tilde{g}}\ln A-\frac{3}{2}\tilde{g}%
^{\mu\nu}\frac{\partial_{\mu}A\partial_{\nu}A}{A^{2}}\right)  .
\end{equation}
Dropping a total divergence and noting that,
$\partial_{\mu}A=\left(  \partial A/\partial\phi^{a}\right)
\partial_{\mu}\phi$, the action transforms into,
\begin{equation*}
\tilde{S}=\int_{\mathcal{M}}dv_{\tilde{g}}\left(
\tilde{R}-\tilde{g}^{\mu \nu }\left(
\frac{3}{2A^{2}}A_{a}A_{b}+\frac{1}{A}h_{ab}\right) \partial _{\mu
}\phi ^{a}\partial _{\nu }\phi ^{b}\right) .
\end{equation*}
In general, the quadratic form
\begin{equation}
\pi_{ab}:=\frac{3}{2A^{2}}A_{a}A_{b}+\frac{1}{A}h_{ab}
\label{quad}
\end{equation}
is not positive definite (unless one imposes further conditions on
$A$)\footnote{For example, in simple scalar-tensor theories with
$A\left(\phi\right) =\phi$ and $B\left( \phi\right)
=\omega\left(\phi\right) /\phi$ one has to impose the condition
that $\omega\left(\phi\right)\geq-3/2$ in order that the energy
density of the scalar field be nonnegative.}. Assuming that
$\mathrm{rank\,}\pi_{ab}=\dim\mathcal{N},$ we may define the
reciprocal tensor $\pi^{ab},$ i.e., $\pi^{ab}\pi_{bc}
=\delta_{c}^{a}$ and endow the target manifold with the new metric
$\pi_{ab}$ and use this metric to raise and lower indices in
$\mathcal{N}.$

Using this metric defined by Eq. (\ref{quad}), the original
wavemap-tensor theory (1)-(2) is conformally equivalent to a
wavemap minimally coupled to general relativity,
\begin{equation}
\tilde{S}=\int_{\mathcal{M}}\tilde{L}_{\sigma}dv_{\tilde{g}},\quad\tilde
{L}_{\sigma}=\sqrt{\tilde{g}}\left(
\tilde{R}-\tilde{g}^{\mu\nu}\pi
_{ab}\partial_{\mu}\phi^{a}\partial_{\nu}\phi^{b}\right)
\label{coaction}.
\end{equation}
This result shows that all couplings of the wavemap to the
curvature are equivalent. Varying this conformally related action,
$\dot{\tilde{S}}=0$, we find the Einstein-wavemap system field
equations for the $\tilde{g}$ metric and involving the $\pi_{ab}$
metric, namely,
\begin{equation}
\tilde{G}_{\mu\nu}=\pi_{ab}\left(
\partial_{\mu}\phi^{a}\partial_{\nu}
\phi^{b}-\frac{1}{2}\widetilde{g}_{\mu\nu}\widetilde{g}^{\rho\sigma}
\partial_{\rho}\phi^{a}\partial_{\sigma}\phi^{b}\right) , \label{einwm}
\end{equation}
\begin{equation}
\tilde{\Box}_{\tilde{g}}\phi^{a}+D_{bc}^{a}\tilde{g}^{\mu\nu}\partial_{\mu
}\phi^{b}\partial_{\nu}\phi^{c}=0,\quad D_{bc}^{a}=\Gamma_{bc}^{a}
(\widetilde{h})+T_{bc}^{a}, \label{emsf}
\end{equation}
where
$T_{abc}:=\partial_{c}Q_{ab}+\partial_{b}Q_{ac}-\partial_{a}Q_{bc}$
and $Q_{ab}:=Q_{a}Q_{b}$ with $Q_{a}=\sqrt{3/2}A^{-1}A_{a}$.

We shall return to  these equations in the last Section where we
shall give a new physical interpretation of their conformal
properties.

\section{Inflation in wavemap-tensor theories}
We now present a way of generating a cosmological constant term
without using a potential. Let us assume that the source manifold
$(\mathcal{M}^{m},g_{\mu\nu})\;$ is a 4-dimensional flat FRW model
in the original  wavemap-tensor theory. The Friedman equation in
the Einstein frame reads,
\begin{equation}
H^{2}=T_{WM}^{00},
\end{equation}
where the 00-component of the wavemap energy-momentum tensor is
given by,
\begin{equation}
T_{WM}^{00}=\pi_{ab}\dot{\phi}^{a}\dot{\phi}^{b}.\label{ooen}
\end{equation}
In this frame, scalar fields evolve according to (\ref{emsf}). It
then follows that at  critical points of $T_{WM}^{00}$ the
universe inflates exponentially. This is the simplest example of a
general procedure, which we call $\sigma$-inflation, in which
inflation is driven both by the coupling $A(\phi)$ \emph{and} the
self-interacting (target manifold is curved!) scalar fields
$(\phi^{a})$ which, however, have no potentials.

This mechanism reduces to the so-called hyperextended inflation
mechanism \cite{stac} when the target space is the real line. On
the other hand, when the curvature coupling $A(\phi)$ is equal to
one, $T_{WM}^{00}$ can have no critical points and so we have no
inflationary solutions. In this case we obtain the so-called
tensor-multiscalar models \cite{dano}. Inflationary solutions
become possible in this case by adding `by hand' extra potential
terms and models of this sort abound.

From (\ref{ooen}) we see that specific forms of the function
$A(\phi)$ lead to a  series of critical points of $T_{WM}^{00}$
corresponding to different values of the cosmological constant.
Further investigation of the stability of these points is expected
to single out a preferred family of functions $A(\phi)$ and target
manifolds $(\mathcal{N}^{n},h_{ab})$ allowing a \emph{slow}
evolution of $T_{WM}^{00}$ (necessary for an adequate amount of
inflation). Hence, large values of the energy density of the
wavemap drive the early universe to an inflationary stage that
lasts until $T_{WM}^{00}$ reaches a critical point leaving us with
a cosmological constant which must be  compatible with
observational parameters and other constraints (see \cite{vile}
for a recent review of the problems associated to the cosmological
constant). This problem is currently under study.

\section{Wavemap-tensor theories as brane worlds}
We showed in Section 3 the  (con)formal equivalence between the
original wavemap-tensor theory (\ref{action}), (\ref{lagran})
 and the  Einstein-wavemap system (\ref{coaction}).
 Suppose now, for the sake of illustration, that
$ \mathrm{dim}\mathcal{N}>\mathrm{dim}\mathcal{M}\;$ so that $\phi
(\mathcal{M})\subset\mathcal{N}$  can be considered as an
$\left(1+(m-1)\right)$--dimensional  subset in the
$\left(1+(m-1)+(n-m)\right)$--dimensional target manifold.

A cursory look at the conformal system (\ref{emsf}) reveals  that
the 'fields' are constrained to propagate only on the `brane' manifold
$(\mathcal{M},\tilde{g})$  since their derivatives are taken only
with respect to the metric $\tilde{g}$. On the other hand, the
gravitational field defined from Eq. (\ref{einwm}) propagates
freely on the `bulk' manifold $(\mathcal{N},\pi_{ab})$ where the
bulk metric is now  the $\pi_{ab}$ metric defined by Eq.
(\ref{quad}). As an example, we can consider a Randall-Sundrum
type model \cite{ra-su99} where the brane
$(\mathcal{M},\tilde{g})$ has minimal codimension ($n-m=1$).

We believe that this is an interesting point which requires
further careful analysis for it provides a new interpretation of
the conformal equivalence result for wavemap-tensor theories.
Hence, a wavemap in the original frame could perhaps be considered
as a matter field there since when conformally transformed becomes
a set of scalar fields which live on the brane
$(\mathcal{M},\tilde{g})$ while gravity propagates off this subset
and into the bulk space $(\mathcal{N},\pi_{ab})$
\footnote{Inclusion of matterfields in the original wavemap-tensor
theory (for instance a perfect fluid) does not seem to alter this
result. In the field equations for the conformally transformed
matter,  derivatives will be again taken only with respect to the
conformal `brane' metric $\tilde{g})$.}.

A more detailed analysis of the results presented here will be
given elsewhere.

\end{document}